\documentclass[preprintnumbers, nofootinbib,eqsecnum,prd, showpacs]{revtex4}

\preprint{PITT-PACC 1210}

\usepackage{graphicx}
\usepackage{amsmath,amssymb}
\usepackage{url}
\usepackage{doi}
\usepackage{color}
\usepackage{multirow}
\usepackage{subfigure}
\usepackage{verbatim}

\newcommand{\lsim}{\mathrel{\mathop{\kern 0pt \rlap
  {\raise.2ex\hbox{$<$}}}
  \lower.9ex\hbox{\kern-.190em $\sim$}}}
\newcommand{\gsim}{\mathrel{\mathop{\kern 0pt \rlap
  {\raise.2ex\hbox{$>$}}}
  \lower.9ex\hbox{\kern-.190em $\sim$}}}
\newcommand{\mev}{{\,{\rm MeV}}}
\newcommand{\gev}{{\,{\rm GeV}}}

\newcommand{\beq}{\begin{equation}}
\newcommand{\eeq}{\end{equation}}
\newcommand{\bea}{\begin{eqnarray}}
\newcommand{\eea}{\end{eqnarray}}

\def\mh{m_{h}}

\def\gmh{\Gamma_h}
\def\mm{\mu^+ \mu^-}
\def\bb{b\bar b}
\def\ww{W W^*}
\def\fbi{{\rm fb}^{-1}}
\def\fb{{\rm fb}}
\def\pb{{\rm pb}}
\def\br{{\rm Br}}
\def\cm2s{{\rm cm^{-2} s^{-1}}}
\begin{document}

\title{Potential precision of a direct measurement of the Higgs boson total width \\ at a muon collider}
\bigskip
\author{Tao Han}
\author{Zhen Liu}
\affiliation{Pittsburgh Particle Physics, Astrophysics and Cosmology Center, Department of Physics and Astronomy,
University of Pittsburgh, 3941 O'Hara St., Pittsburgh, Pennsylvania 15260, USA}
\date{\today}
\begin{abstract}
In the light of the discovery of a $126~\gev$ Standard-Model-like Higgs boson at the LHC, we evaluate the achievable accuracies for direct measurements of the width, mass, and the $s$-channel resonant production cross section of the Higgs boson at a proposed muon collider.
We find that with a beam energy resolution of $R=0.01\%~(0.003\%)$ and integrated luminosity of $0.5~\fbi~(1~\fbi)$, a muon collider would enable us to determine the Standard-Model-like Higgs width
to $\pm0.35~\mev\ (\pm0.15~\mev)$ by combining two complementary channels of the $WW^*$ and $b\bar b$ final states.
A non-Standard-Model Higgs with a broader width is also studied. The unparalleled accuracy potentially attainable at a muon collider would test the Higgs interactions to a high precision.

\end{abstract}
\maketitle

\section{Introduction}
The discovery of a Higgs boson at the LHC \cite{ATLASHnew,CMSHnew} completes the simple structure of the Standard Model (SM). Yet, a profound question remains: Is this rather light, weakly-coupled boson nothing but a SM Higgs, or it is a first manifestation of a deeper theory? While the LHC certainly will take us to a long journey on seek for new physics beyond the SM, it would be very important to determine the Higgs boson's properties as accurately as possible at the LHC and future collider facilities, whether or not there are other particles directly associated with the Higgs sector observed at the LHC.

Of all properties of the Higgs boson, its total decay width ($\gmh$) is perhaps of the most fundamental importance since it characterizes the overall coupling strength. Once it is determined, the partial decay widths to other observable channels would be readily available.
Because of the broad spread of the partonic energy distribution, limited energy-momentum resolution for final-state particles and the large SM backgrounds in the LHC environment,
there is essentially no way to measure its total width or any partial width to a desirable accuracy without additional theoretical assumptions \cite{Duhrssen:2004cv,Couple}. Assuming an upper limit for a Higgs coupling, such as that of $hWW$, then an upper bound for the total width can be inferred \cite{Peskin}.
At an International Linear Collider (ILC) optimized for Higgs boson studies,
the $ h ZZ$ coupling and thus the partial decay width $\Gamma(h\to ZZ)$ can be measured to a good accuracy. The total decay width then may be indirectly determined to $6\%-11\%$ \cite{ILC}.
At a muon collider, however, due to the much stronger coupling for the Higgs to the muons than to the electrons, an $s$-channel production of a Higgs boson \cite{Barger:1995hr}
will likely lead to clear signal for several channels, and thus its total decay width may be directly measured by fitting its scanned data.

In this paper, we propose a realistic scanning and fitting procedure to determine the Higgs boson width at a muon collider. We demonstrate the complementarity for the two leading signal channels $h\to b\bar b,\ WW^{*}$. The combined results lead to a highly accurate determination for the width, mass and the $s$-channel production cross section. This is undoubtedly invaluable for determining the Higgs interactions and testing
the theory of the electroweak symmetry breaking to an unparalleled precision.

\section{Resonant Profile for a Higgs Boson}
For a resonant production $\mm \to h$ and a subsequent decay to a final state $X$ with a collider c.m.~energy
$\sqrt{\hat s}$,  the Breit-Wigner formula reads
\bea
\sigma(\mm \to h \to X)
= \frac{4\pi \Gamma_h^{2} \br(h\to \mm )\br(h\to X )} {(\hat s-m^2_h)^2 + \Gamma_h^2 m_h^2},
\label{eq:sigma}
\eea
where $\br$ denotes the corresponding decay branching fraction.
At a given energy, the cross section is governed by three parameters: $m_h$ for the signal peak position, $\Gamma_h$ for  the line shape profile, and the product $B \equiv {\br}(h\to \mu^+\mu^- ){\br}(h\to X )$ for the event rate.

In reality, the observable cross section is given by the convolution of the energy distribution delivered by the collider. Assume that the $\mm$ collider c.m.~energy ($\sqrt s$) has a flux distribution
$$ { dL(\sqrt s) \over d\sqrt{\hat s} }
= {\frac 1 {\sqrt{2\pi \Delta}} } \exp[\frac {-( \sqrt{\hat s} - \sqrt s)^2} {2\Delta^2}] ,
$$
with a Gaussian energy spread $\Delta  = R \sqrt {s}/\sqrt{2}$, where $R$ is the percentage beam energy resolution; then, the effective cross section is
\bea
\nonumber
&& \sigma_{\rm eff}(s) = \int d \sqrt{\hat s}\  \frac {dL(\sqrt s)} {d\sqrt{\hat s}}   \sigma(\mm \to h \to X) \\
&& \propto  \left\{
\begin{array}{ll}
\Gamma_h^{2} B / [( s-m^2_h)^2 + \Gamma_h^2 m_h^2]
\quad~   (\Delta  \ll \Gamma_{h}), & \\
B \exp[{ \frac {-( \mh - \sqrt s)^2 } {2\Delta^2} }]
(\frac {\Gamma_h}  {\Delta}) / m^{2}_h
  \quad  (\Delta \gg \Gamma_{h}).  &
\end{array}
\right.
\label{eq:convol}
\eea
For  $\Delta  \ll \Gamma_{h}$, the line shape of a Breit-Wigner resonance can be mapped out by scanning over the energy as given in the first equation.
For $\Delta  \gg \gmh$ on the other hand, the physical line shape is smeared out by the Gaussian distribution of the beam energy spread, and the signal rate will be determined by the overlap of the Breit-Wigner and the luminosity distributions, as seen in the second equation above.

Unless stated otherwise, we focus on the SM Higgs boson with the mass and total width as
\bea
\mh=126~\gev,\quad \gmh=4.21\ \mev .
\eea
For definitiveness in this study,
we assume two sets of representative values for the machine parameters \cite{muC}
\bea
 {\rm Case\ A:} && R=0.01\% \ (\Delta=8.9\mev), \ L=0.5~\fbi, ~~~~~
\label{eq:a} \\
 {\rm Case\ B:} && R=0.003\% \ (\Delta=2.7\mev), \  L=1~\fbi.
\label{eq:b}
\eea
We see that their corresponding beam energy spread $\Delta$ is comparable to the Higgs total width.
In Fig.~\ref{tab:cpara}, we show the effective cross section versus the $\mm$ collider c.m.~energy for the SM Higgs boson production. A pure Breit-Wigner resonance is shown by the dotted curve. The solid and dashed curves include the convolution of the luminosity distribution for the two beam energy resolutions and are integrated over $\sqrt {\hat s}$.
For simplicity, we have taken the branching fractions $h\to \mu^+\mu^-$ to be the SM value and the final state $h\to X$ to be $100\%$. The beam energy resolution manifests its great importance in comparison between the solid and dashed curves in this figure.

\begin{figure}[tb]
\includegraphics[width=180pt]{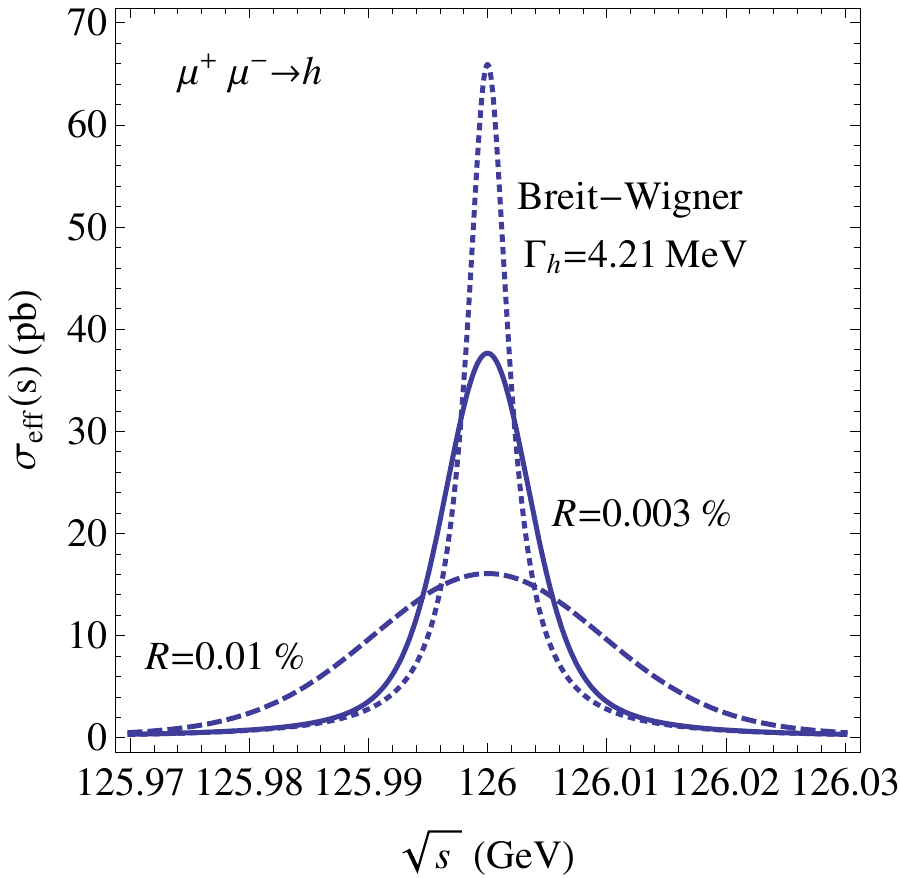}
\caption{Effective cross section for $\mm \to h$ versus the collider energy $\sqrt s$ for the SM Higgs boson production with $\mh=126~\gev$. A Breit-Wigner line shape with $\gmh=4.21$ MeV is shown (dotted curve). The solid and dashed curves compare the two beam energy resolutions of cases A and B.}
\label{fig:xs}
\end{figure}

\section{Width Determination for the SM Higgs Boson}

\begin{table}[tb]
\centering
\begin{tabular}{|c|c|c|c|c|c|}
  \hline
 & $\mm\rightarrow h$ & \multicolumn{2}{|c|}{$h\rightarrow b\bar b$} &  \multicolumn{2}{|c|}{$h\rightarrow WW^*$}  \\ \cline{3-6}
 \raisebox{1.6ex}[0pt]{R (\%)} & $\sigma_{\rm eff}$ (pb) & $\sigma_{Sig}$ & $\sigma_{Bkg}$ &  $\sigma_{Sig}$ & $\sigma_{Bkg}$ \\ \hline
 $0.01$ & $16$ & $7.6$ & & $3.7$ &  \\ \cline{1-3} \cline{5-5}
$0.003$ & $38$ & $18$ & \raisebox{1.6ex}[0pt]{$15$}  & $5.5$ & \raisebox{1.6ex}[0pt]{$0.051$}\\ \hline
\end{tabular}
\caption{Effective cross sections (in pb) at the resonance $\sqrt {s}=\mh$ for two choices of beam energy resolutions $R$ and two leading decay channels, with the SM branching fractions $\br_{b\bar b}=56\%$ and $\br_{WW^*}=23\%$~\cite{Handbook}.}
\label{tab:cpara}
\end{table}

An excellent beam energy resolution for a muon collider would make a direct determination of the Higgs boson width possible in contrast to the situations in the LHC and ILC. Because of the expected narrow width for a SM Higgs boson, one still needs to convolute the idealistic Breit-Wigner resonance with the realistic beam energy spectrum as illustrated in Eq.~(\ref{eq:convol}). We first calculate the effective cross sections at the peak for the two cases of energy resolutions A and B.
We further evaluate the signal and SM background for the leading channels
\bea
h \to b \bar b,\quad WW^{*}.
\label{eq:chan}
\eea

We impose a polar angle acceptance  for the final-state particles,
\bea
10^\circ < \theta  < 170^\circ.
\eea
Tightening up the polar angle to $20^\circ-160^{\circ}$ will further reduce the signal by $4.6\%$, and the background by $6.7\%~(15\%)$ for the $b\bar b~(WW^*)$ final states.
We assume a $60\%$ single $b$-tagging efficiency and require at least one tagged $b$ jet for the $b\bar b$ final state.
The backgrounds are assumed to be flat with cross sections evaluated right at $126~\gev$ using Madgraph5 \cite{MG5}. This appears to be an excellent approximation over the energy range of the current interest about 100 MeV.
We tabulate the results in Table \ref{tab:cpara}.
The background rate of $\mm\rightarrow Z^*/\gamma^* \to b\bar b$ is $15~\pb$, and the rate of $\mm \to WW^*\to 4$ fermions is only $51~\fb$, as shown in Table~\ref{tab:cpara}. Here, we consider all the decay modes of $WW^*$ because of its clear signature at a muon collider.
The four-fermion backgrounds from $Z\gamma^*$ and $\gamma^*\gamma^*$ are smaller to begin with and can be greatly reduced by kinematical considerations such as by requiring the invariant mass of one pair of jets to be near $m_W$ and setting a lower cut for the invariant mass of the other pair.
While the $b\bar b$ final state has a larger signal rate than that for $WW^*$ by about a factor of three, the latter has a much improved signal (S) to background (B) ratio, about 100:1 near the peak.

For a given beam resolution, we assume that a scan procedure over the collider
c.m.~energy $\sqrt s$ is available. The current Higgs mass statistical error is about 0.4 GeV \cite{ATLASHnew,CMSHnew} with an integrated  luminosity of about 10 $\fbi$.
Toward the end of the LHC run with about 100 times more luminosity accumulated, it is conceivable to improve the statistical error of the mass determination by about an order of magnitude. Then, the systematic errors would have to be controlled to a best level. It was argued that an ILC could reach a similar or better accuracy \cite{Barger:1996pv}. We thus proceed to scan over the energy in the range
\bea
126\gev \pm 30\mev\ \ {\rm in\ 20\ scanning\ steps.}
\label{eq:scan}
\eea
The energy scanning step is set at $3\mev$, roughly the same size of the $\Delta$ and $\Gamma_h$.

We first generate ideal data in accordance with a Breit-Wigner resonance at this mass convoluted with Gaussian distribution of the beam energy integrated over $\sqrt {\hat s}$, as discussed before.
These data are then randomized with a Gaussian fluctuation with standard deviation $\sqrt N$, where $N$ is the number of events expected for a given integrated luminosity, summing both signal and background. The simulated events over the scanning points are plotted with statistical errors for the assumed integrated luminosity as in Eqs.~(\ref{eq:a}) and (\ref{eq:b}).

\begin{figure}[bt]
\begin{center}
\subfigure{
      \includegraphics[width=120pt]{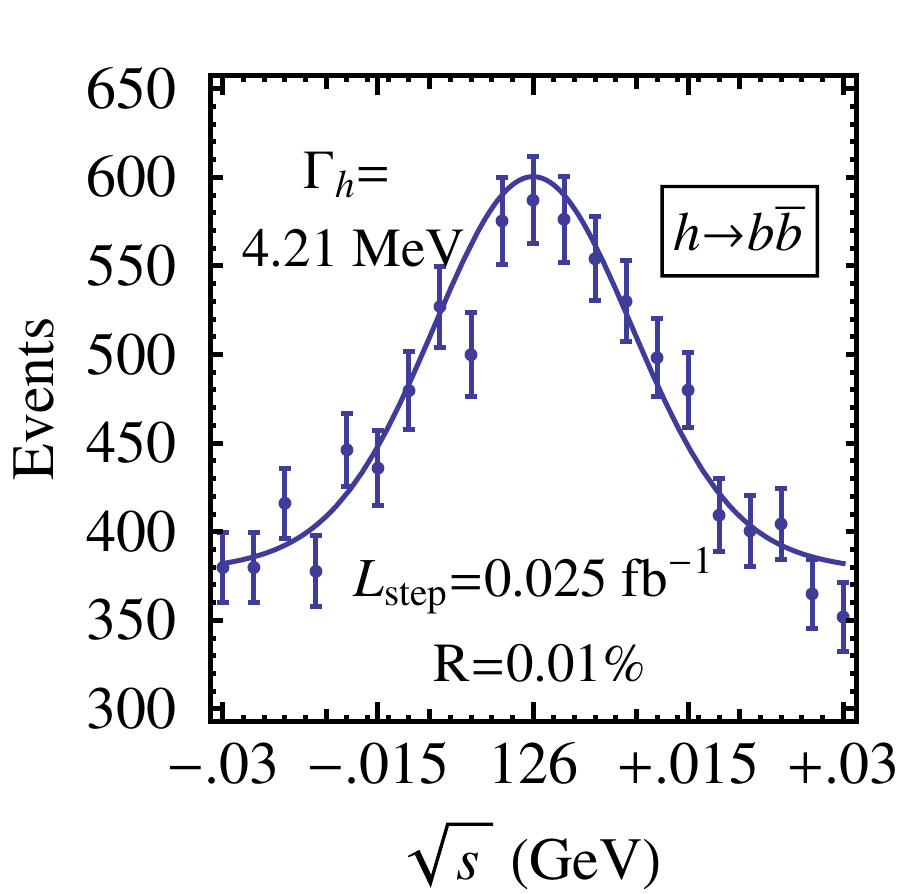}}
\subfigure{
      \includegraphics[width=120pt]{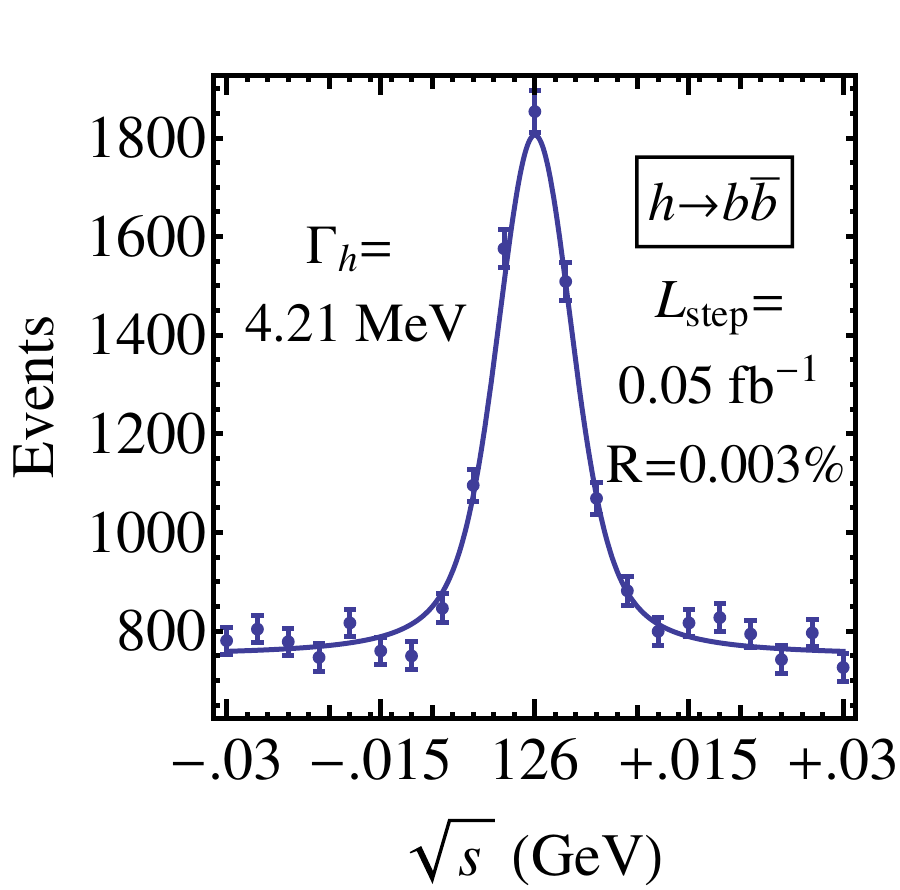}}
\subfigure{
      \includegraphics[width=120pt]{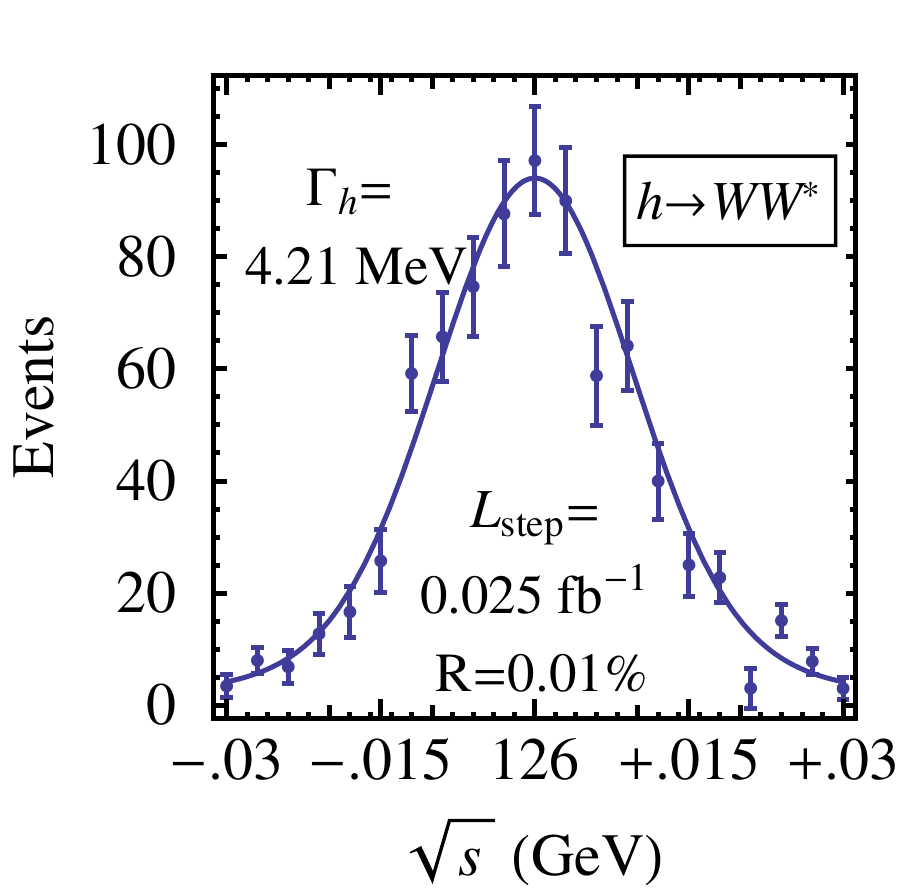}}
\subfigure{
      \includegraphics[width=120pt]{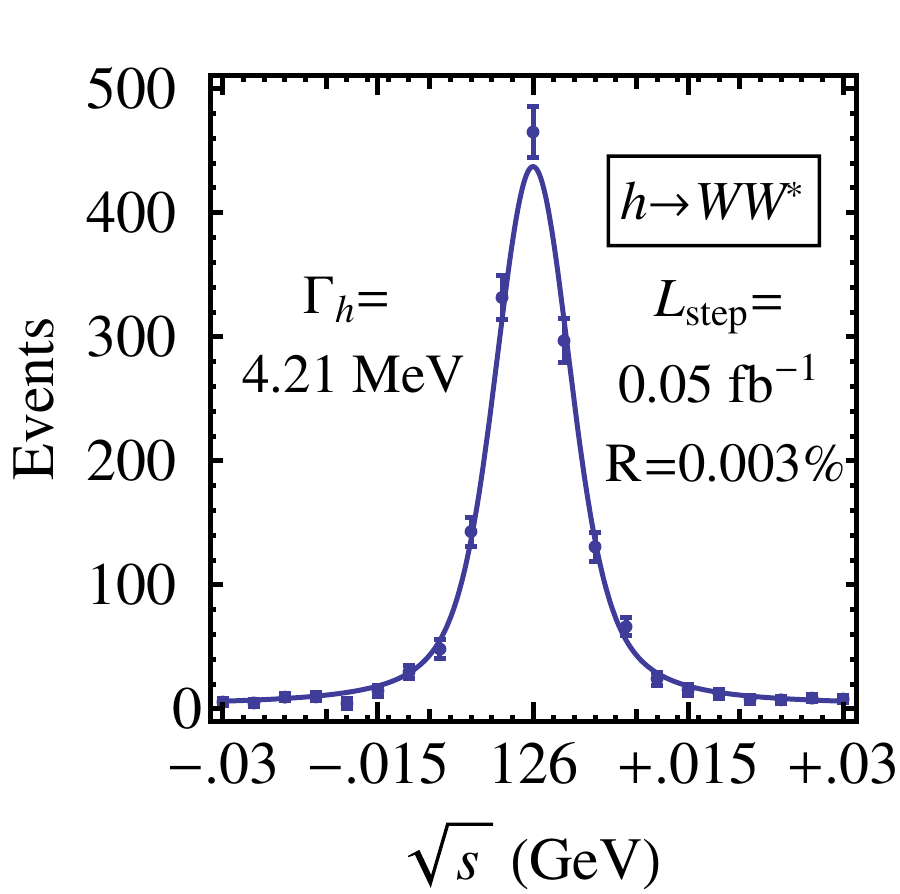}}
\caption{Number of events of the Higgs signal plus backgrounds and statistical errors expected for cases A and B as a function of the collider energy $\sqrt s$ in $b\bar b$ and $WW^*$ final states with a SM Higgs $\mh =126~\gev$ and $\gmh = 4.21~\mev$.}
\label{fig:shape12}
\end{center}
\end{figure}

The results are shown by the solid curves in Fig.~\ref{fig:shape12}, for cases A and B as in Eq.~(\ref{eq:a}) (left panels) and Eq.~(\ref{eq:b}) (right panels). The $\bb$ and $\ww$ final states are separately shown by the upper and lower panels, respectively.

\begin{figure}[b]
\begin{center}
\subfigure{
      \includegraphics[width=145pt]{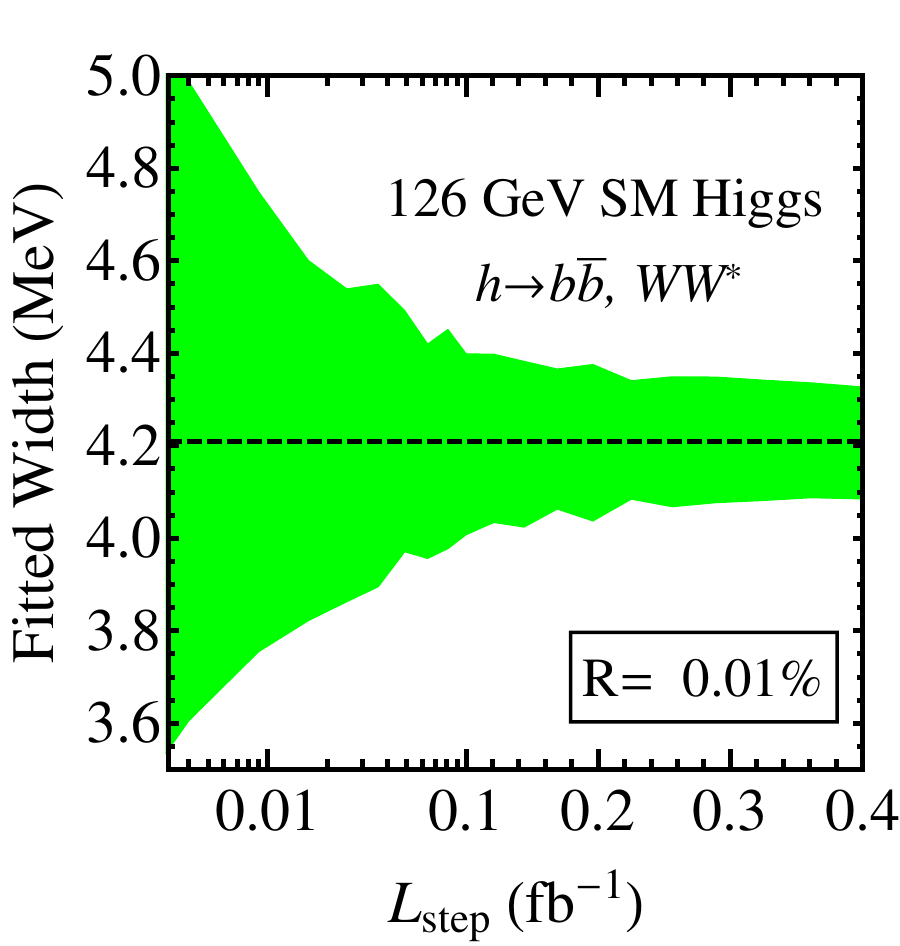}}
\subfigure{
      \includegraphics[width=145pt]{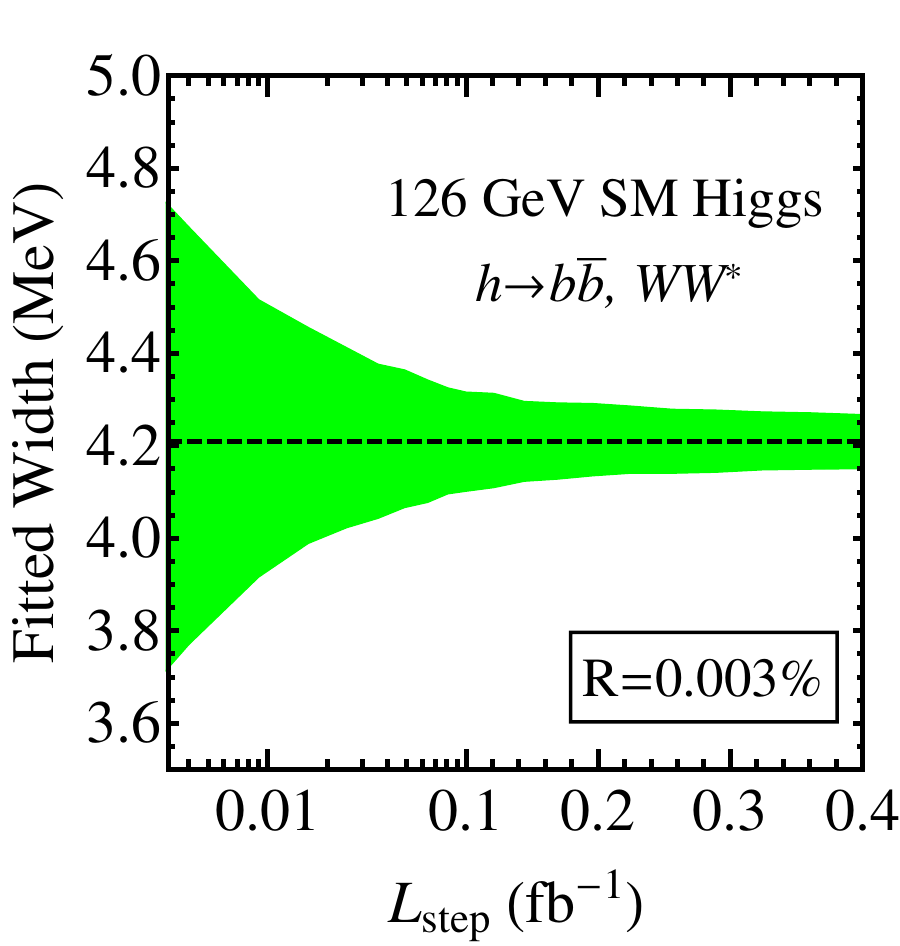}}
\caption{Fitted values and errors for the SM Higgs width versus the luminosity per step with the scanning scheme as specified in Eq.~(\ref{eq:scan}).
}
\label{fig:luminofit}
\end{center}
\end{figure}

We adopt a $\chi^{2}$ fit over the scanning points with three model-independent free parameters in the theory
$\gmh,\ B$ and $\mh$ by minimizing the $\chi^2$.
The fitting accuracies for the Higgs properties can be illustrated by the standard deviation, denoted  by $\delta \gmh$, $\delta B$, and $\delta m_h$. These standard deviations are estimated by the standard method of projecting the $\chi^2_{min}+1$ sphere to corresponding parameters.
To see the effects from the available luminosity, we show our results for the SM Higgs width determination in
Fig.~\ref{fig:luminofit} for both cases by varying the luminosity.
The achievable accuracies with the scanning scheme as specified in Eq.~(\ref{eq:scan})
by combining two leading channels are summarized in Table~\ref{tab:acrcy} for three representative luminosities per step with the same 20-step scanning scheme.

\begin{table}[tb]
\begin{tabular}{|c|c|c|c|c|}
  \hline
  $\Gamma_h=4.21~\mev$ & $L_{\rm step}$ ($\fbi$) & $\delta{\Gamma_h}~(\mev)$ & $\delta B$ & $\delta m_{h}~(\mev)$ \\ \hline \hline
  & $0.005$ & $0.73$ & $6.5\%$  & $0.25$ \\ \cline{2-5}
  $R=0.01\%$ & {\bf 0.025} & {\bf 0.35} & {\bf 3.0\%} & {\bf 0.12} \\  \cline{2-5}
  & $0.2$ & $0.17$ & $1.1\%$ & $0.06$ \\ \hline \hline
  & $0.01$ & $0.30$ & $4.4\%$ & $0.12$ \\ \cline{2-5}
  $R=0.003\%$ & {\bf 0.05} & {\bf 0.15} & {\bf 2.0\%} & {\bf 0.06} \\ \cline{2-5}
  & $0.2$ & $0.08$ & $1.0\%$ & $0.03$ \\
  \hline
\end{tabular}
\caption{Fitting accuracies for one standard deviation of $\Gamma_h$, $B$, and $m_h$ of the SM Higgs with the scanning scheme as specified in Eq.~(\ref{eq:scan}) for three representative luminosities per step. Results with the default luminosities for cases A and B described in Eqs.~(\ref{eq:a}) and (\ref{eq:b}) are in boldface.}
\label{tab:acrcy}
\end{table}


\section{Width Determination for a Broader Higgs Boson}

We now explore the unique feature of the direct width measurement for a broader resonance at a muon collider.
For definitiveness, we still work with a Higgs-like particle with a mass at 126 GeV, but with a total width of ten times larger than that of the SM value, $\Gamma_{h}=42$ MeV. We shall consider scenarios in which the signal at the LHC of this particle (assuming a SM Higgs) would be unchanged.

\begin{table}[b]
\centering
\begin{tabular}{|c|c|c|c|c|c|}
  \hline
 & $\mm\rightarrow h$ & \multicolumn{2}{|c|}{$h\rightarrow b\bar b$} &  \multicolumn{2}{|c|}{$h\rightarrow WW^*$}  \\  \cline{3-6}
 \raisebox{1.6ex}[0pt]{R (\%)} & $\sigma_{\rm eff}$ (pb) & $\sigma_{Sig}$ & $\sigma_{Bkg}$ &  $\sigma_{Sig}$ & $\sigma_{Bkg}$ \\ \hline
 $0.01$ & $18$ & $2.6$ & & $1.3$ & \\ \cline{1-3} \cline{5-5}
 $0.003$ & $20$ & $3.0$ & \raisebox{1.6ex}[0pt]{$15$}  & $1.5$ & \raisebox{1.6ex}[0pt]{$0.051$} \\ \hline
\end{tabular}
\caption{The effective cross sections (in pb) for the exotic Higgs, with $\br_{b\bar b}=18\%$ and $\br_{WW^*}=7.3\%$.}
\label{tab:cparabroader}
\end{table}

\begin{figure}[tb]
\begin{center}
\subfigure{
      \includegraphics[width=120pt]{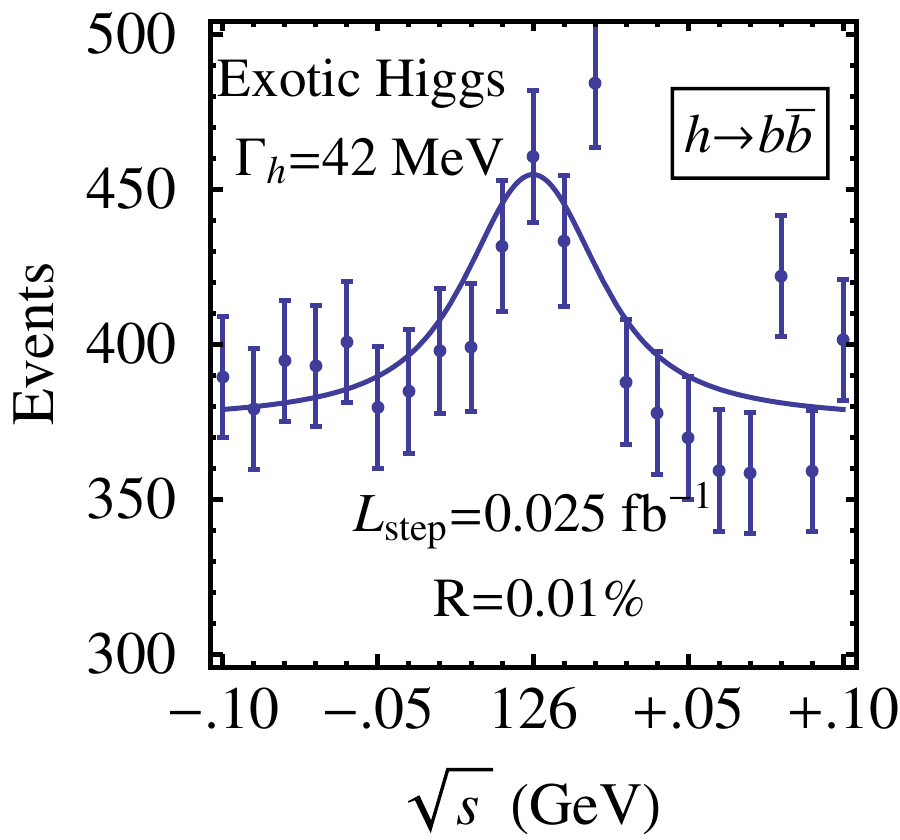}}
\subfigure{
      \includegraphics[width=120pt]{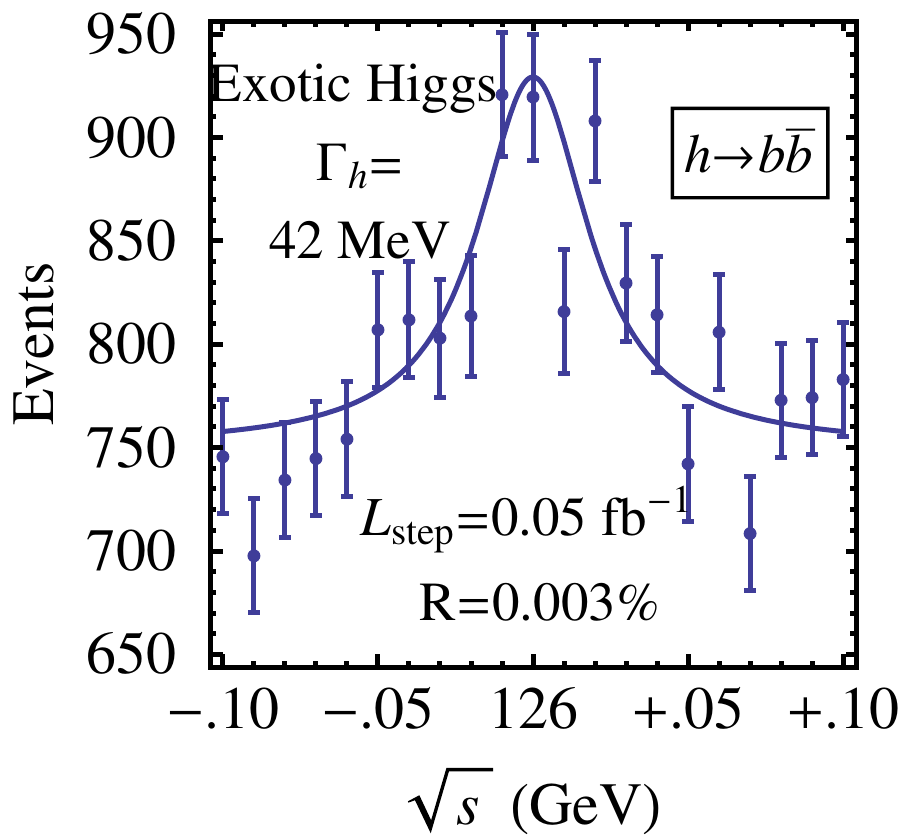}}
      \subfigure{
      \includegraphics[width=120pt]{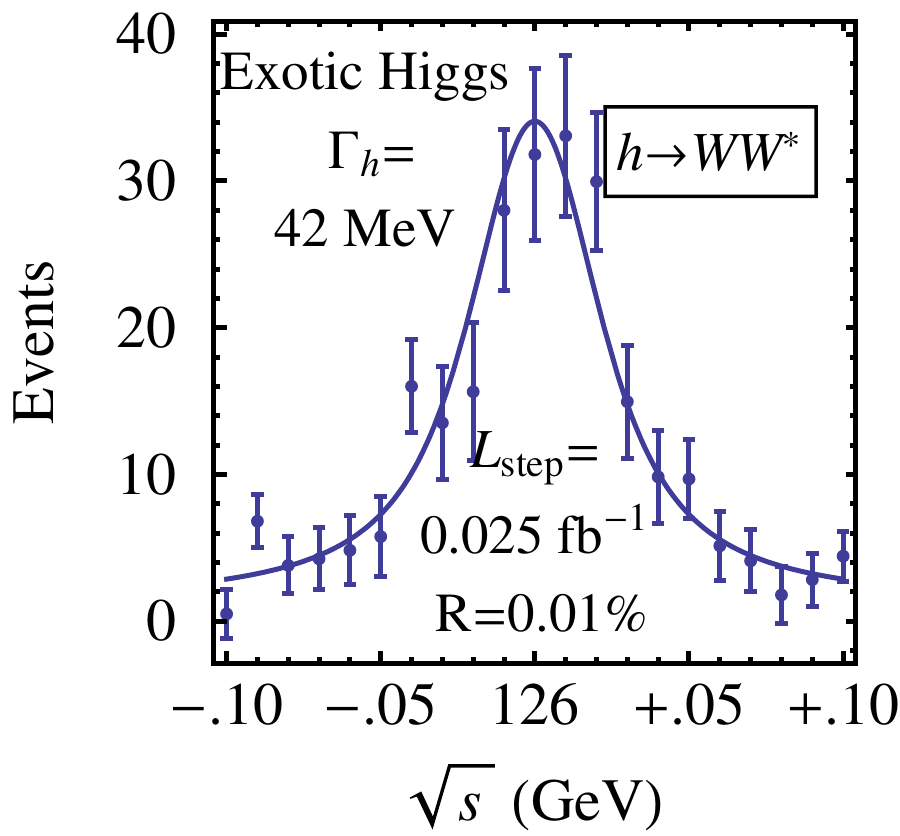}}
\subfigure{
      \includegraphics[width=120pt]{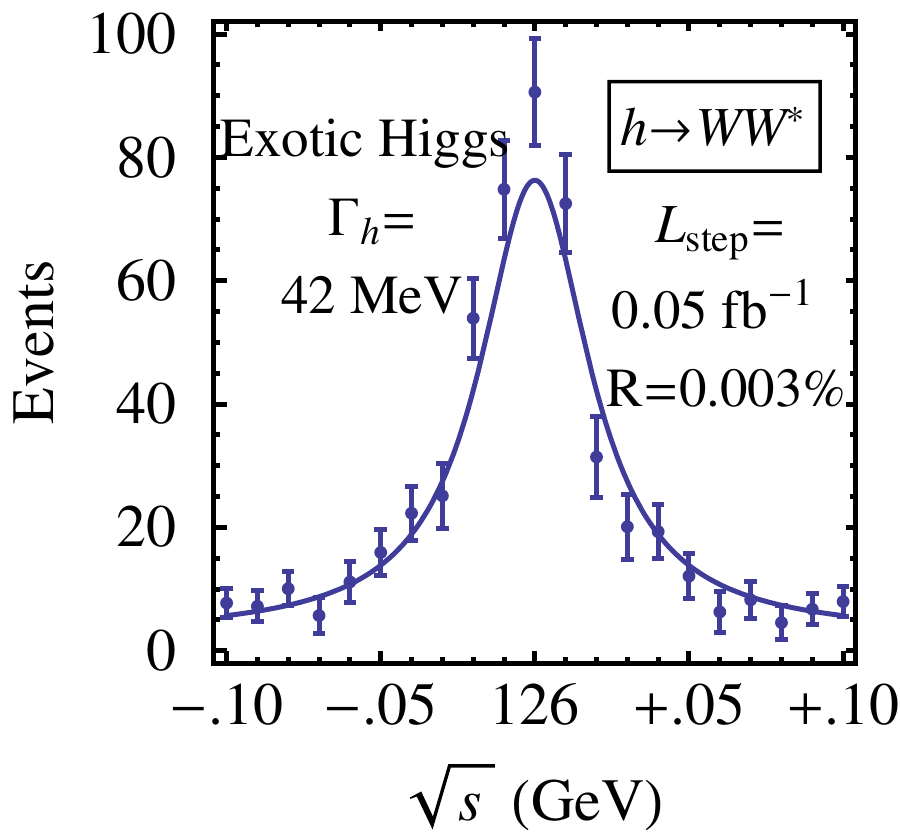}}
\caption{Number of events of the Higgs signal plus backgrounds
and statistical errors expected for cases A and B as a function of the collider energy $\sqrt s$ in $b\bar b$ and $WW^*$ final states with an exotic Higgs $\mh =126~\gev$ and $\gmh = 42~\mev$.}
\label{fig:shape12fat}
\end{center}
\end{figure}

In Fig.~\ref{fig:shape12fat}, we present the similar analyses as in Fig.~\ref{fig:shape12} for a broader Higgs.
There are two features of this figure compared to the SM Higgs in Fig.~\ref{fig:shape12}. First, the increase of Higgs width requires a broader scan range to reconstruct the Breit-Wigner resonant distribution. We choose to scan the same number of 20 scan steps with a step size of $10~\mev$, while keeping the same total integrated luminosity. It is seen from the figure that the physical line shape of the Higgs boson is essentially mapped out by the scanning. Second, since the signal rate at the LHC is governed by partial widths to initial ($i$) and final ($f$) states $\propto \Gamma_{i} \Gamma_{f} /\Gamma_{h}$, the rate could be kept the same when increasing the Higgs total width by a factor $\kappa$ while scaling the partial widths up by a factor of $\sqrt{\kappa}$. This would correspondingly reduce the cross section for the signal at a muon collider as seen in Eq.~(\ref{eq:sigma}).
Under this constraint, the results for the branching fractions and the effective peak cross sections of a broader Higgs at a muon collider are listed in Table~\ref{tab:cparabroader}.

\begin{figure}[tb]
\begin{center}
\subfigure{
      \includegraphics[width=145pt]{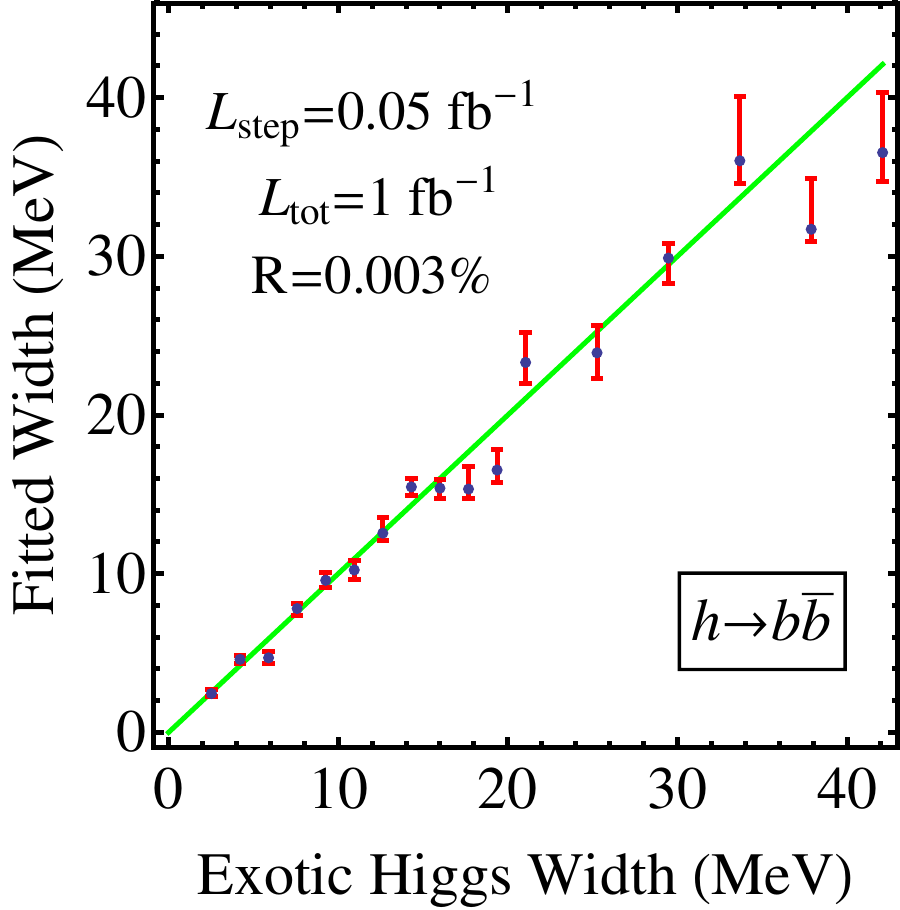}}
\subfigure{
      \includegraphics[width=145pt]{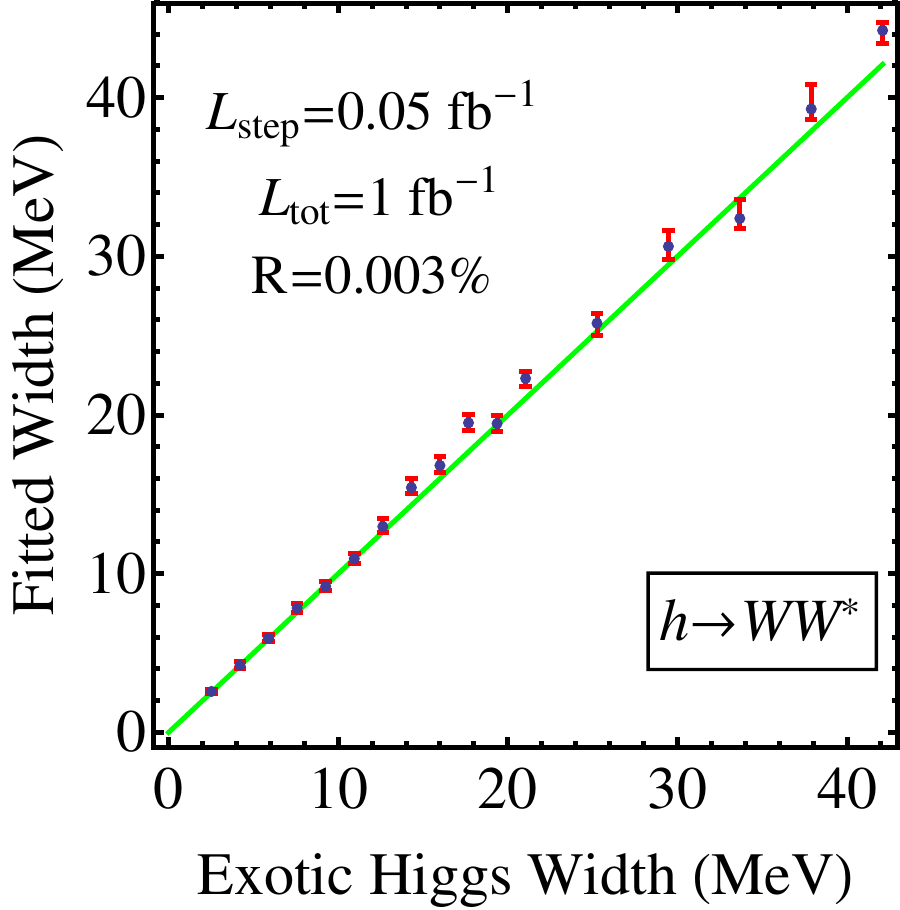}}
\caption{Fitted values and errors for the Higgs width versus the input values. The step size is set as a rounded half-integer value between $3~\mev-10~\mev$ in accordance with the Higgs width $0.6-10$ times the SM value.}
\label{fig:widthfit}
\end{center}
\end{figure}

Although a larger Higgs width would be easier to resolve with a fine energy resolution, it is a practical concern when a larger range of the scanning energy is needed with a fixed total luminosity.
In Fig.~\ref{fig:widthfit}, we explore this issue by plotting the width determination with statistical errors for a fixed total luminosity at $1~\fbi$ and varying Higgs widths. The events from the $b\bar b$ channel and $WW^*$ channel are shown individually.
It turns out that a smaller width could receive better accuracies in the scanning process due to the larger signal rate than that at a larger width as mentioned above. We summarize the fitting accuracies in Table~\ref{tab:widthscan}.
\begin{table}[tb]
\centering
\begin{tabular}{|c|c|c|c|}
  \hline
  $\gmh = 2.5-42$ MeV & $\delta\Gamma_h~(\mev)$ & $\delta B$ & $\delta{m_h}~(\mev)$ \\ \hline
  $b\bar b$ & $0.33-1.1$ & $2.8\%-11\%$ & $0.07-1.2$ \\ \hline
  $WW^*$ & $0.14-0.67$ & $2.2\%-5.2\%$ & $0.05-0.71$  \\
  \hline
\end{tabular}
\caption{Fitting accuracies for the exotic Higgs properties. The scanning scheme is the same as in Fig.~\ref{fig:widthfit}. }
\label{tab:widthscan}
\end{table}

\section{Discussions and Conclusions}
The direct measurement of the Higgs width with high precision will be invaluable to explore new physics through this ``Higgs lamp post."
For instance, varying the parameters $\tan\beta,\ M_{A}$ in the minimal supersymmetric standard model (MSSM) within the current LHC constraints, the SM-like Higgs width could change by $20\%$ \cite{MSSM}.
Models with Higgs invisible decays would increase the width.
Generic Higgs multiplet models allow an increase in total width, as illustrated in the triplet Georgi-Machacek model \cite{Georgi}.
Composite Higgs models also alter the Higgs width from the SM value.

The mass and cross section can be simultaneously determined along with the Higgs width to a high precision.
The results obtained are largely free from theoretical uncertainties.
Uncertainties of the signal evaluation do not alter the width and mass fitting. The major systematic uncertainty comes from our knowledge of beam properties \cite{muC}. The uncertainty associated with the beam energy resolution $R$ will directly add to our statistical uncertainties of Higgs width. This uncertainty can be well-calibrated by experimentalists as well as by measuring the Z boson peak rate. Our estimated accuracies are by and large free from detector resolutions. Other uncertainties associated with b tagging, acceptance, etc., will enter into our estimation of signal strength $B$ directly. These uncertainties will affect our estimation of total width $\Gamma_h$ indirectly through statistics, leaving a minimal impact in most cases. It is worthy it to mention that our scanning scheme for the SM Higgs case is by simply adopting the projected accuracy for the Higgs mass measurement from the LHC and ILC. A prescanning with the muon collider to narrow down the mass window could also increase the achievable accuracies, as a tradeoff for the total luminosity available.

Moreover, our study on the width and mass measurements can be applicable to new particles predicted in many theories. For example, the $CP$-odd and the other $CP$-even Higgs states in the minimal supersymmetric standard model and in two-Higgs-doublet-models may all be suitably studied at a muon collider. The achievable high accuracy would help to resolve nearly degenerate Higgs states.

In conclusion, the newly observed Higgs-like particle at the LHC strongly motivates a muon collider as the Higgs factory. We proposed methods and evaluated the attainable accuracy to directly measure the Higgs width by scanning and fitting the $s$-channel resonance. The unparalleled precision would test the Higgs interactions to a high precision
and undoubtedly take us to a deeper understanding of the electroweak-symmetry-breaking sector.

\begin{acknowledgments}
\vskip -0.5cm
We thank Estia Eichten, Ron Lipton, and David Neuffer for discussions.
This work is supported in part by the U.S.~Department of Energy under Grant No. DE-FG02-95ER40896, in part by PITT PACC. Z.L.~is in part supported by the LHC Theory Initiative from the U.S. National Science Foundation under Grant No. NSF-PHY-0969510. T.H.~would also like to thank the Aspen Center for Physics for the hospitality during which part of the work was carried out. A.C.P. is supported by the NSF under Grant No.  1066293.

\end{acknowledgments}


\end{document}